\documentclass[aps,pra,reprint,superscriptaddress]{revtex4-2}

\usepackage{graphicx}% Include figure files
\usepackage{bm}% bold math
\usepackage{soul,xcolor}
\usepackage{physics}
\usepackage{amsmath}
\usepackage{siunitx}
\usepackage{tabularx}
\usepackage{float}
\usepackage{booktabs}
\usepackage{array,multirow}
\usepackage{hyperref}
\usepackage{orcidlink}

\hypersetup{
    colorlinks=true,
    linkcolor=blue,
    citecolor=blue,
    filecolor=blue,
    urlcolor=blue
}

\newcolumntype{Y}{>{\raggedright\arraybackslash}X}

\begin{document}

\preprint{APS/123-QED}

\title{Continuous cloud position spectroscopy using a magneto-optical trap}

\author{Benedikt Heizenreder\,\orcidlink{0000-0001-5611-4144}}
\thanks{These authors contributed equally to this work.}
\affiliation{Van der Waals-Zeeman Institute, Institute of Physics, University of Amsterdam, Science Park 904, 1098 XH Amsterdam, The Netherlands}

\author{Ananya Sitaram\,\orcidlink{0000-0002-6548-4515}}
\thanks{These authors contributed equally to this work.}
\affiliation{Van der Waals-Zeeman Institute, Institute of Physics, University of Amsterdam, Science Park 904, 1098 XH Amsterdam, The Netherlands}

\author{Sana Boughdachi\,\orcidlink{0009-0004-6305-2438}}
\affiliation{Van der Waals-Zeeman Institute, Institute of Physics, University of Amsterdam, Science Park 904, 1098 XH Amsterdam, The Netherlands}
\affiliation{Toptica Photonics SE, Lochhamer Schlag 19, 82166 Gr\"afelfing, Germany}

\author{Andrew von H\"orsten \,\orcidlink{0009-0005-5048-3877}}
\affiliation{Toptica Photonics SE, Lochhamer Schlag 19, 82166 Gr\"afelfing, Germany}

\author{Yan Xie\,\orcidlink{0009-0000-8349-2888}}
\affiliation{VSL National Metrology Institute, Thijssweg 11, 2629 JA Delft, The Netherlands}

\author{Andreas Brodschelm\,\orcidlink{0009-0000-4293-2031}}
\affiliation{Toptica Photonics SE, Lochhamer Schlag 19, 82166 Gr\"afelfing, Germany}

\author{Florian Schreck\,\orcidlink{0000-0001-8225-8803}}
\email{CloudPositionSpectroscopy@strontiumBEC.com} 
\affiliation{Van der Waals-Zeeman Institute, Institute of Physics, University of Amsterdam, Science Park 904, 1098 XH Amsterdam, The Netherlands}
\affiliation{QuSoft, Science Park 123, 1098 XG Amsterdam, The Netherlands}

\date{\today}

\date{\today}
%%%%%%%%%%%%%%%%%%%%%%%%%%%%%%%%%%%%%%%%%%%%%%%%%%%%%%%%%%%%%%%%%%%%%%%%%%%%%%%%%%%%%%%%%%%%%
\begin{abstract}
We demonstrate a continuous spectroscopy technique with frequency sensitivity well below the natural transition linewidth, while maintaining a locking range hundreds of times larger. The method exploits the position dependence of a continuous, broadband magneto-optical trap operating on the 7.5\,kHz-wide intercombination line of strontium. We show that the frequency sensitivity is fundamentally insensitive to the effective MOT laser linewidth. By applying active feedback on the MOT position to a dispersion-optimized frequency comb, which serves as the reference for stabilizing the MOT laser~\cite{boughdachi2025strontium}, we achieve a frequency instability below $4.4\times10^{-13}$ after 400\,s of averaging in both the optical and radio-frequency domains, surpassing the performance of conventional hot-vapor modulation transfer spectroscopy. Our method is a broadly applicable alternative route to frequency references in the high $10^{-14}$ range around 100\,s.
\end{abstract}

\maketitle            
%%%%%%%%%%%%%%%%%%%%%%%%%%%%%%%%%%%%%%%%%%%%%%%%%%%%%%%%%%%%%%%%%%%%%%%%%%%%%%%%%%%%%%%%%%%%%
Continuously operating atomic clocks, such as hydrogen masers in the RF domain~\cite{Polyakov_2021,LAURENT2015}, and systems employing modulation transfer spectroscopy in the optical domain~\cite{McCarron2008,Miao2022,Lee2023,Doeringshoff2019}, have demonstrated frequency stability on the order of $\sim 10^{-14}$ after 100~s of averaging~\cite{McCarron2008,Miao2022,Lee2023,Doeringshoff2019,dierikx_white_2016}.
Recent advances have enabled the development of compact, field-deployable optical atomic clocks, including systems based on tetrahedral and pyramidal magneto-optical traps (MOTs)~\cite{Nosske2025Transportable,Sitaram2020,Grating_MOT_Bondza2022}, microcell-based clocks~\cite{tsygankov_single-frequency_2025,klinger_cs_2025}, and vapor-based clocks~\cite{Shelving_spectroscopy_of_the_strontium_intercombination_line, Hilton2025,guan_velocity-comb_2025}.
While modulation transfer spectroscopy remains widely used in these systems, it is limited by short-term probe laser instability and a narrow frequency locking range, both typically on the order of the transition linewidth~\cite{preuschoff2018linewidth,zhang2018frequency}. These limitations are especially pronounced for narrow-line optical transitions, such as those found in the alkaline-earth and alkaline-earth-like atoms. Additionally, modulation transfer spectroscopy does not provide direct access to the RF domain.

In this Letter, we present a simple, continuous spectroscopy technique that exploits the dependence of the MOT position along the vertical axis (gravity) on the MOT laser frequency in a five-beam configuration. We demonstrate that this method improves both the frequency locking range and tolerance to short-term probe laser instability by two orders of magnitude compared to the natural linewidth, while simultaneously achieving a spectral resolution more than an order of magnitude below the 7.5\,kHz natural linewidth of the $\mathrm{^{88}Sr}$ strontium intercombination transition.
By applying active feedback based on the MOT position to a dispersion-optimized frequency comb, used for stabilizing the MOT laser, we extract both an 800\,MHz RF reference and an optical reference at 689\,nm, reducing frequency instabilities below $4.4 \times 10^{-13}$ after 400\,s of averaging. Finally, we demonstrate that this technique can outperform modulation transfer spectroscopy with hot atomic vapors in terms of averaging times of $\sim 100~\mathrm{s}$ on the same transition.

\textit{Theory}.\label{sec:theory} To model the vertical dynamics of the five-beam MOT, where in comparison to a standard MOT the sixth beam propagating in gravity direction is removed~\cite{boughdachi2025strontium}, we consider the balance between gravitational force \(mg\) and optical force along the vertical axis for a single atom with mass \(m\)~\cite{metcalf1999laser}. The resulting steady-state condition determines the MOT position as a function of laser detuning and magnetic field gradient~\cite{Loftus_line_cooling,MOT_pos_Ytterbium_Sillus}.
For strontium, using the \(\rm ^1S_0 \rightarrow\, ^3P_1\) intercombination line at 689~nm with a narrow linewidth of \(\gamma_0 = 2\pi \cdot 7.5\,\mathrm{kHz}\), power broadening alone is insufficient to achieve large capture velocities. Instead, frequency modulation is used to extend the effective capture range by more than an order of magnitude. While this ``broadband (BB) MOT" technique increases the capture velocities, it cannot be modeled simply as a power-broadened single-frequency (SF) light force.
In our implementation, the RF drive of the AOM is triangularly modulated between $\delta_{\text{start}}$ and $\delta_{\text{stop}} = \delta_{\text{start}} - 2\pi f_{\text{bandwidth}}$, where both $\delta$ values denote detunings from resonance, and $f_{\text{bandwidth}}$ is the total modulation range~\cite{Bennetts2019SteadyStateStrontium,boughdachi2025strontium}. In the low-saturation regime (see Appendix~\ref{App:Saturation Limit}), the total force in the vertical direction is:
\begin{equation}
    F_{\text{total}} = mg - \frac{\hbar k \gamma_0 I_{\text{total}}}{2I_0} \sum_{i=0}^{n} \frac{\alpha_i}{1 + \left(2\delta_i/\gamma_0\right)^2},
\label{eq:broad_modulated}
\end{equation}
where
\begin{equation}
   \delta_i = \delta_{\text{start}} - i 2\pi f_{\text{mod}} \mp kv - \frac{\mu_B g_J B(z)}{\hbar},
\label{eq:broad_MOT_frequency}
\end{equation}
where $n = \mathrm{round}\!\left( f_{\text{bandwidth}} / f_{\text{mod}} \right)$ is the total number of frequency components considered, \(f_{\text{mod}}\) is the modulation frequency, \(\sum_{i=0}^{n}\alpha_i=1\) are the relative intensities of the \(i^\text{th}\) frequency component, \(I_{\text{total}}=2P/(\pi\omega_0^2)\) is the total intensity of the MOT beam (defined by the beam power \(P\) and gaussian waist 
size \(\omega_0\)) and \(I_0\) the saturation intensity.
For \( m_J = \pm 1\) of this transition, the Zeeman shift is \(\frac{\mu_B g_J}{h} =2101~\text{kHz/G}\)~\cite{Bennetts2019SteadyStateStrontium}, where \(\mu_B\) is the Bohr magneton and \(g_J\) is the Landé g-factor. In equilibrium, \(F_{\text{total}} = 0\), and the MOT position is directly related to the Zeeman shift. 
Since we operate close to the center of a quadrupole magnetic field with a linear gradient \(B'\), a change \(\delta_{\text{start}} \rightarrow \delta_{\text{start}} + 2\pi \Delta f\) results in a corresponding vertical displacement \(\Delta z\), where the frequency-to-position conversion factor \(\Delta f / \Delta z=\frac{\mu_B g_J B'}{\hbar}\) is independent of natural linewidth \(\gamma_0\) and power broadening.
%%%%%%%%%%%%%%%%%%%%%%%%%%%%%%%%%%%%%%%%%%%%%%%%%%%%%%%%%%%%%%%%%%%%%%%%%%%%%%
\begin{figure}[]
\centering
\includegraphics[width=8.3cm]{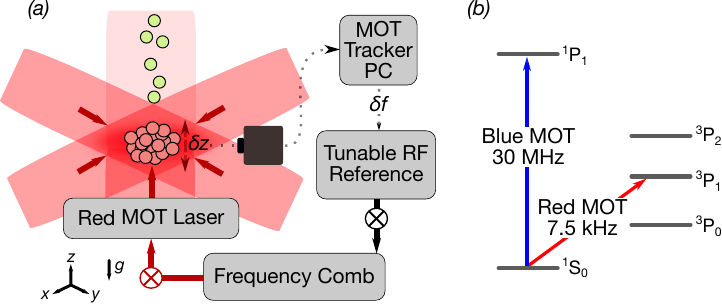}
\caption[Experimental Setup]{
Experimental schematic:  
(a) Atoms (green disks) are continuously loaded from an upper 2D blue MOT ((b), operating on the \(\rm ^1S_0 \rightarrow ^1P_1\) transition at 461~nm) into a five-beam broadband $\mathrm{^{88}Sr}$ red MOT ((b), operating on the narrow \(\rm ^1S_0 \rightarrow ^3P_1\) transition at 689~nm). The red MOT is monitored non-destructively via fluorescence imaging, and the center position along the vertical (gravity) axis is extracted every 50\,ms.
The red MOT laser is locked to an optical frequency comb, which provides short-term frequency stability, while long-term stability is achieved via the MOT position where the vertical displacement \(\delta z\) serves as an error signal. Feedback is applied through a tunable RF oscillator used to synchronize the comb, which in turn transfers its stability to both the RF and optical domains.}
\label{fig:Setup}
\end{figure}

\begin{figure}[]
   \centering
    \includegraphics[width=8.3cm]{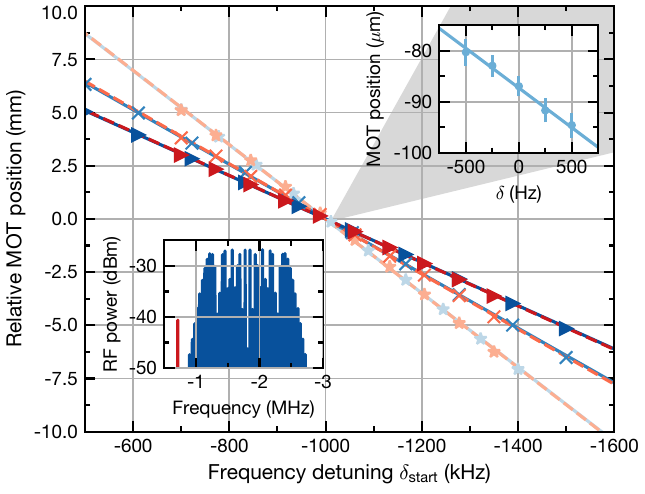}
    \caption[RF]{Frequency sensitivity of the broadband (BB) and single-frequency (SF) MOTs, demonstrating MOT laser linewidth independence. Despite differing forces, both MOTs exhibit similar sensitivities within error bars. Triangles: vertical magnetic field gradient \(0.47\,\mathrm{G/cm}\), sensitivities \(98.19(30)\,\mathrm{Hz/\mu m}\) (BB, blue) and \(98.76(50)\,\mathrm{Hz/\mu m}\) (SF, red). Crosses: \(0.37\,\mathrm{G/cm}\), with \(77.91(30)\,\mathrm{Hz/\mu m}\) (light blue, BB) and \(77.21(40)\,\mathrm{Hz/\mu m}\) (light red, SF). Stars: \(0.27\,\mathrm{G/cm}\), with \(57.37(24)\,\mathrm{Hz/\mu m}\) (light orange, BB) and \(57.25(40)\,\mathrm{Hz/\mu m}\) (turquoise, SF). An offset magnetic field shifts the intersection point of the six curves to \(\delta_{\text{start}}=1.05\,\mathrm{kHz}\). Bottom inset: MOT AOM RF spectra showing BB modulation (blue) and SF (red) tone.} Top inset: high-resolution BB MOT position scan near the intersection point, with frequency resolution below \(\num{250}\,\mathrm{Hz}\) at \(0.31\,\mathrm{G/cm}\).
    \label{fig:multi_vs_SF}
\end{figure}

\textit{Experimental Setup}.\label{sec:methods} 
We employ our previously developed architecture for generating a narrow-line $\mathrm{^{88}Sr}$ MOT using a dispersion-optimized frequency comb-locked laser system, without the need for a free space high-finesse optical cavity~\cite{boughdachi2025strontium}.
In contrast to typical experiments that execute cooling stages time-sequentially, we execute them in a spatial sequence. 
Atoms effuse from an oven, are Zeeman slowed and then cooled in a blue 2D MOT, out of which they continuously fall downwards, forming a stream of millikelvin-temperature atoms.
These atoms are subsequently captured by a second-stage cooling system in a separate chamber: a five-beam, red BB MOT operating on the narrow \(\rm ^1S_0 \rightarrow\,^3P_1\) transition [see Fig.~\ref{fig:Setup}(a)]. The red MOT quadrupole field is generated by a coil pair of approximately 20\,cm diameter, arranged in an anti-Helmholtz configuration with vertical axis.
The MOT laser beams are frequency modulated by a single-pass AOM, which is driven by a modulated RF signal. This signal is generated by a DDS (Analog Devices AD9852) with a triangular frequency modulation with a total modulation range of $f_{\text{bandwidth}} = 1.5~\mathrm{MHz}$ and modulation frequency $f_{\text{mod}} = 40~\mathrm{kHz}$ (see Fig.~\ref{fig:multi_vs_SF}, inset). Further details on the apparatus and methodology can be found in Ref.~\cite{boughdachi2025strontium}.

To characterize the MOT position dependence on the red MOT laser frequency, as well as shifts caused by other perturbations, we use absorption imaging, which is inherently destructive. 
For these initial calibration measurements, the red MOT laser source (a Toptica DL pro followed by an injection-locked diode laser) is frequency-stabilized in a conventional way: on short timescales ($<1$\,s) to an optical cavity and on longer timescales to a modulation transfer spectroscopy signal from a hot Sr vapor~\cite{StellmerPhD2013,qiao_ultrastable_2019}. For comb and conventional system, the MOT laser linewidth is below the natural transition linewidth \(\gamma_0 \).

Although our apparatus is designed for continuous operation, we can also generate a pulsed single frequency MOT (SF MOT), increasing phase space density and reducing the cloud temperature well below a microkelvin \cite{boughdachi2025strontium}. 
In this procedure, we first load the BB MOT with several hundred million atoms, then transfer them to an SF MOT by switching to a single frequency tone while keeping the magnetic field unchanged. 
After a 100\,ms equilibration period that allows the MOT to stabilize, we take an absorption image. 
To compare the frequency sensitivity (i.e., the frequency-to-position conversion factor) of the SF MOT to that of the BB MOT, the sum in Eq.~\ref{eq:broad_modulated} reduces to the $i=0$ term for the SF MOT.

When using the BB MOT position as a frequency reference, we lock the MOT laser source to the frequency comb (\ref{App:Frequency Transfer}). The comb repetition rate is synchronized to a tunable, low-noise 800\,MHz oscillator (TOPTICA RF TUNE),which serves as a short-term reference, while long-term frequency stability is transferred via the MOT position to the tunable RF oscillator (\ref{App:RF-Tuning}).
To enable continuous feedback, we switch to non-destructive fluorescence imaging. The MOT position is extracted every 50\,ms via live fitting of the fluorescence image. Using the known frequency-to-position conversion factor \(\Delta f / \Delta z\), a Python-based proportional-integral (PI) loop continuously feeds back to the tunable RF oscillator.
This stabilization scheme, utilizing the comb, effectively transfers frequency stability to both the RF and optical domains simultaneously. We monitor both using a  K+K counter (FXE Phase + Frequency Meter); the RF is mixed down and measured directly, while the optical stability is measured via a beat note between the DL Pro (locked to the MOT position) and our conventional laser system. The K+K counter is synchronized to a stable 10\,MHz RF reference from the Dutch Metrology Institute (VSL), which has an fractional instability of approximately \(10^{-14}\) after 100 seconds of averaging~\cite{dierikx_white_2016,VanTour2017WhiteRabbit, surf2024timefrequency,boughdachi2025strontium}.

\textit{Sensitivity and linewidth independence}.\label{sec:sensitivity} We first characterize the sensitivity of both MOT types' (BB and SF) position to changes in the laser frequency using the conventional laser system.
Specifically, we measure the relationship between the MOT laser detuning \(\delta_{\text{start}}\) and the MOT position along the vertical direction, shown in Fig.~\ref{fig:multi_vs_SF}. 
We measure over the wide frequency range for which we can form a stable MOT, from approximately 600\,kHz to 1.5\,MHz red-detuned from the transition, at three different vertical magnetic field gradients ($B' < 1$\,G/cm). 
For all magnetic field gradients, we observe a linear relationship between the MOT position and the frequency detuning \(\delta_{\text{start}}\) for the BB MOT and the SF MOT. A linear fit to the data using the Zeeman shift allows us to extract the magnetic field gradient experienced by the atoms. For the smallest gradient (0.27\,G/cm), we measure a sensitivity, or frequency-to-position conversion factor \(\Delta f / \Delta z\), for the BB MOT (orange stars) of 57.37(24)\,Hz/$\mu$m and for the SF MOT (light blue stars) of 57.25(40)\,Hz/$\mu$m. 
We are able to resolve frequency shifts as small as 250\,Hz by restricting the scan range around the intersection point of the three sensitivity curves, where magnetic field gradient fluctuations have the least effect (see inset top of Fig.~\ref{fig:multi_vs_SF}). 
This sensitivity is more than 30 times smaller than the natural linewidth of the transition, indicating the potential to outperform conventional modulation transfer spectroscopy, specifically considering narrow intercombination lines as they can be found in alkaline-earth(-like) atoms.
Notably, the usable locking range of our system corresponds to the frequency range over which a stable BB MOT can be maintained, spanning approximately 1\,MHz. 
This is two orders of magnitude wider than the typical locking range achievable with conventional transfer modulation spectroscopy on the same transition, which is usually limited to a few \(\gamma_0\)~\cite{preuschoff2018linewidth}.

As determined in the theory section, the frequency-to-position conversion factor is insensitive of the MOT laser linewidth, as long as a five-beam MOT can be created. 
The BB MOT laser has a linewidth that is broadened by more than two orders of magnitude, due to modulation and power broadening, whereas in the SF MOT, operated in the low-intensity regime (\(I \approx I_{\text{sat}}\)), the effective linewidth is expected to remain close to the natural linewidth (see inset bottom of Fig.~\ref{fig:multi_vs_SF}). 
We observe that the frequency-to-position conversion factor agrees within experimental uncertainties across all three magnetic field gradients for both MOT types, see Fig.~\ref{fig:multi_vs_SF}. 
This supports the conclusion that MOT position spectroscopy reduces the requirements for the MOT/probing laser linewidth as long as it is constant over time. 

\begin{table}[b]
\centering
\begin{tabular}{|p{5.25cm}|p{1.8cm}|}
\hline
\textbf{Noise Source} & \textbf{Stability}  \\\hline
Power (vertical direction) & $\pm 0.5\%$  \\\hline
Power (horizontal direction) & $\pm 0.1\%$  \\\hline
Atom number & $\pm 0.9\%$  \\\hline
Vertical magnetic field & $\pm 33.4~\mu $G  \\\hline
Horizontal magnetic\,\,field & $\pm 340~\mu $G  \\\hline
Vertical magnetic quadrupole field & $\pm 3.9~$mG/cm \\\hline
\end{tabular}
\caption{Summary of dominant noise sources limiting long-term MOT position stability at a magnetic field gradient of \(0.39\,\mathrm{G/cm}\) around \(\delta_{\text{start}}=1.05\,\mathrm{kHz}\). The listed stability values represent the required control levels to maintain a position uncertainty of \(\pm 1\,\mu\mathrm{m}\) averaged over 100~s, corresponding to a fractional frequency instability of \(\sigma(100\,\mathrm{s}) = 1.9 \times 10^{-13}\). Details of the individual noise characterizations are provided in Appendix~\ref{App:Shifts Measurements}.}
\label{table:ErrorBudget}
\end{table}

\textit{Systematic errors}.\label{sec:Systematic_errors} Before we can use MOT position spectroscopy as a long-term frequency reference, we must evaluate potential systematic errors that could limit its stability. 
We begin with an error budget analysis at a magnetic field gradient of 0.39\,G/cm with an observed frequency-to-position conversion factor of 82.5(10)\,Hz/$\mu$m, a good balance between frequency sensitivity and magnetic field gradient strength. The parameters that contribute most significantly to uncertainty in the MOT position are: fluctuations in the optical powers of the vertical and horizontal MOT beams, atom number fluctuations, drifts in the background magnetic field, and instabilities in the quadrupole magnetic field gradient.
Magnetic field variations are the dominant contribution to the error budget, as they directly affect the local Zeeman shift and thus lead to displacements of the MOT’s equilibrium position. The effects of atom number fluctuations and beam intensity variations are more subtle. 
A detailed characterization of the sensitivity of the MOT position to each of these parameters is provided in Appendix~\ref{App:Shifts Measurements}.
The required stability levels for each of these parameters to limit the MOT position uncertainty to \(\pm1~\mu\)m after 100\,s of averaging are summarized in Table~\ref{table:ErrorBudget}. This level of spatial control corresponds to a fractional frequency instability of below \(1.9\times10^{-13}\), which should be well within reach using a combination of active and passive stabilization~\cite{Sohmen2023,Active_stabilization_of_Borkowski}.
Note that in the following implementation, we actively compensate only for power fluctuations; for all other effects, we rely on the passive stability of the system.
%=========================================================
\begin{figure}[]
    \centering
    \includegraphics[width=8.3cm]{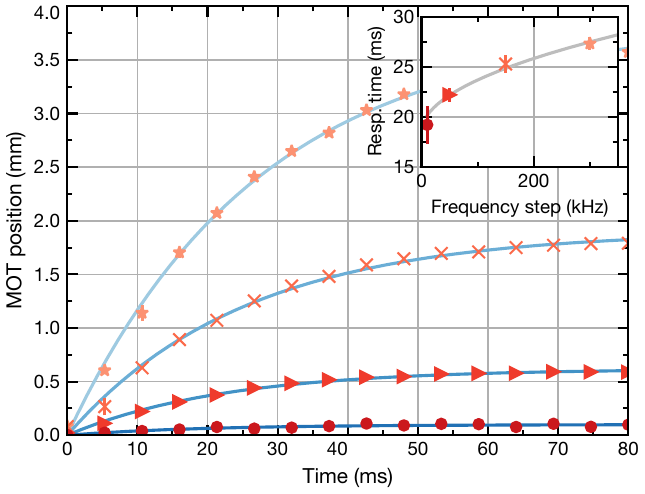}
    \caption[MOT Response Time]{Measurement of the MOT response time following a step change in the red MOT laser frequency. The MOT position relaxes to a new equilibrium according to an overdamped harmonic oscillator model. (Red disks): \(10~\text{kHz}\) jump; (light red triangles): \(50~\text{kHz}\); (orange crosses): \(150~\text{kHz}\); (light orange stars): \(300~\text{kHz}\). 
    Inset: Extracted response time \(\tau\) as a function of frequency jump, fit with a square-root dependence resulting in an amplitude \(A = 0.529(40)~\text{s/kHz}^{1/2}\) and a offset \(\tau_{\text{offset}} = 18.49(57)~\text{ms}\). This relationship sets an upper bound on the MOT response time for a given frequency change.}
    \label{fig:MOT_response_time}
\end{figure}
%=========================================================

\textit{Response Time}.\label{sec:Response_Time} To evaluate the achievable feedback bandwidth for continuous measurement of the MOT position, it is essential to characterize the system’s response time i.e., the time required for the MOT to reach a new equilibrium after a change in MOT laser frequency. Figure~\ref{fig:MOT_response_time} shows the BB MOT position as a function of time following frequency jumps of 10~kHz (red disks), 50~kHz (light red triangles), 150~kHz (orange crosses), and 300~kHz (light orange stars), using the same parameters we later employ in the long-term stability measurement (averaging times of $\sim 100~\mathrm{s}$). 
The MOT's motion can be described by an overdamped harmonic oscillator model, with the position evolving as \( z(t) = z_{\text{start}} + \left(1 - e^{-t/\tau}\right)(z_{\text{end}} - z_{\text{start}}) \), where \( z_{\text{start}} \) and \( z_{\text{end}} \) denote the initial and final MOT positions, and \( \tau \) is the characteristic response time. As shown in the inset of Fig.~\ref{fig:MOT_response_time}, \( \tau \) increases with the magnitude of the frequency jump. We model this dependence empirically as \( \tau(f) = A \sqrt{f} + \tau_{\text{offset}} \). For a frequency change of less then 100~kHz, the system reaches equilibrium in about 25~ms, corresponding to a maximum feedback bandwidth of roughly 40~Hz.
\begin{figure}[]
    \centering
    \includegraphics[width=8.3cm]{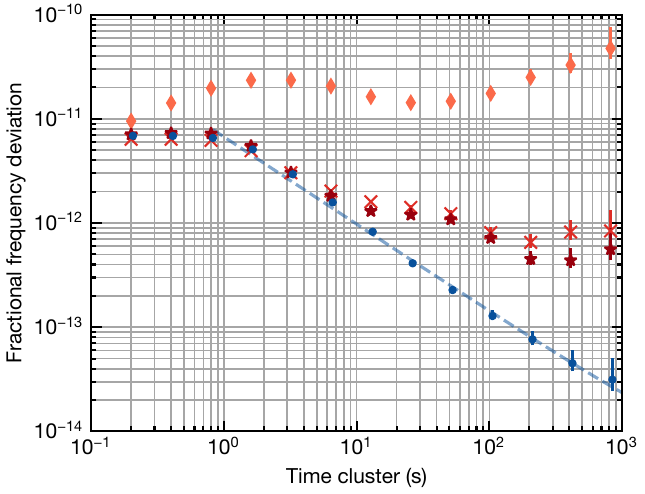}
\caption[Allan Variance RF]{Overlapping ADEV of the free-running tunable RF oscillator (orange diamonds) and the same oscillator stabilized to the MOT position (red stars). Beat signal between the MOT-referenced and hot-vapor-locked lasers (red crosses). Between \(100\)–\(1000~\text{s}\), the MOT position locked RF oscillator can outperform the optical reference based on the hot vapor modulation transfer spectroscopy, reaching a minimum instability of \(4.4 \times 10^{-13}\) at \(\sim400~\text{s}\), limited mainly by magnetic field fluctuations. All RF signals are compared to a stable 10\,MHz RF reference from the Dutch Metrology Institute (VSL). The in-loop error signal (blue points) is fit by \(\sigma(\tau) = (6.71 \pm 0.48) \times 10^{-12} / \tau^{0.84 \pm 0.03}\) (blue dashed line). Error bars indicate $1\sigma$ confidence intervals using a chi-squared distribution.}
    \label{fig:allan_variance_RF}
\end{figure}

\textit{Continuous MOT position spectroscopy}.\label{sec:MOT position spectroscopy} Finally, we implement continuous MOT position spectroscopy using fluorescence imaging of the BB MOT (10\,ms exposure time, 20\,Hz repetition rate). The red MOT laser is stabilized to a frequency comb, and the feedback loop is closed by transferring the MOT position stability to the tunable RF oscillator, which provides the frequency reference for the comb.
We compare the overlapping Allan deviation (ADEV with $100\,\mathrm{ms}$ gate time) for a total measurement time of $T = 4600\,\mathrm{s}$. Error bars indicate $1\sigma$ confidence intervals, calculated using a chi-squared distribution with conservative $T/\tau - 1$ degrees of freedom where $\tau$ is the cluster time~\cite{Riley2008}. Comparing the free-running RF oscillator to the MOT position stabilized case, we observe an improvement of almost two orders of magnitude after 400\,s of averaging (see Fig.~\ref{fig:allan_variance_RF}).The system reaches a minimum fractional frequency instability of $4.4 \times 10^{-13}$ at 400\,s, mainly limited by magnetic field fluctuations.
Notably, we see that the MOT position stabilized tunable RF oscillator can outperform the beat signal between the MOT laser (locked via the comb to the MOT position) and the conventional laser system in the range of 100 to 1000\,s. This points to a limitation of the conventional laser system due to the hot vapor spectroscopy. These results demonstrate that MOT position spectroscopy can exceed the frequency stability of hot vapor modulation transfer spectroscopy for narrow intercombination lines in alkaline-earth(-like) atoms for averaging times of $\sim 100~\mathrm{s}$ with the potential for even longer time scales, even when relying solely on passive stability for magnetic field and atom number. With the addition of active stabilization, the system could approach the performance limit set by the in-loop error signal, calculated by converting MOT positions (recorded with feedback on) into frequency space via the frequency-to-position conversion factor. Therefore, the Allan deviation is expected to follow \(\sigma(\tau) = (6.71 \pm 0.48) \times 10^{-12} / \tau^{0.84 \pm 0.03},\) potentially reaching the high \(10^{-14}\) range after just over 100\,s of averaging (see Appendix~\ref{App:Shifts Measurements}), where we also expect the noise floor to be limited by magnetic field fluctuations.
%%%%%%%%%%%%%%%%%%%%%%%%%%%%%%%%%%%%%%%%%%%%%%%%%%%%%%%%%%%%%%%%%%

\textit{Summary and outlook}.\label{sec:Summary and outlook} 
We have demonstrated, to the best of our knowledge, the first continuous spectroscopy based on the position of a narrow-line strontium MOT, achieving a frequency sensitivity 30 times narrower than the natural linewidth of the transition. This method provides frequency stability surpassing that of hot vapor modulation transfer spectroscopy for averaging times of $\sim 100~\mathrm{s}$ with the potential for even longer time scales, while maintaining a locking range and effective laser linewidth two orders of magnitude broader than the transition's natural linewidth.
If we actively stabilize the MOT position by feeding back to the repetition rate of a dispersion-optimized frequency comb~\cite{boughdachi2025strontium}, we reach a minimum fractional instability of less than $7\times10^{-13}$ after 400\,s of averaging in both the RF and optical domains, primarily limited by magnetic field fluctuations. With magnetic shielding~\cite{Sohmen2023} or active stabilization of stray environmental fields~\cite{Active_stabilization_of_Borkowski}, it is possible to reduce the uncertainty even closer to the limit set by the in-loop error signal.
Further improvement of the short-term stability could be achieved by synchronizing the tunable RF oscillator to a lower noise quartz oscillator~\cite{rakon_hso14} or a four-quadrant photodiode for higher-bandwidth MOT position measurements.
On the other hand, one could also leverage the magnetic field sensitivity of this spectroscopy method and use such a system as a magnetometer in the direction of gravity~\cite{Vengalattore_Magnetometry_2007}. 

This technique can be applied to any element for which a low-temperature five-beam MOT can be achieved (Appendix~\ref{App:Temperature Limit}), such as ytterbium~\cite{karim2025}, erbium, and dysprosium~\cite{Ilzhofer2018}, or for which the final MOT temperature is sufficiently low that the contribution of the sixth beam propagating in the direction of gravity can be neglected~\cite{MOT_pos_Ytterbium_Sillus,chaneliere_three_2008}.
The spectroscopic sensitivity of the MOT position is only dependent on the magnetic field gradient required for the MOT and the transition-specific Land\'e g-factor, meaning that any cycling transition of any species for which these parameters are suitable can be used to convert MOT position to laser frequency with high resolution~(\ref{App:Temperature Limit}).

The use of the frequency comb makes this method convenient, robust, and versatile, while also enabling transfer of the stability into the RF domain, making this method well-suited for applications that require long-term stable RF references, such as Global Navigation Satellite System redundancy~\cite{petrov2016distributed,hasan2018gnss}. In the future, the scheme could be simplified even more by using an injection-lock-amplified comb tooth, which is tuned directly to the target transition. 
Nevertheless, MOT position spectroscopy does not require a frequency comb or a continuous MOT. The stability of a pulsed MOT could also be transferred to a tunable cavity, which is commonly used in cold atom experiments. Here, the MOT cycle time needs to be fast enough to keep the MOT inside the locking range.
This simple, widely-applicable spectroscopy method opens up a host of possibilities for advancing quantum technologies.
%=====================================================
\begin{acknowledgments}
We acknowledge helpful discussions with M. Borkowski and E. Baaij. We thank SURF for maintaining the White Rabbit link between VSL and UvA. We also thank R.J.C. Spreeuw, Mikhail Volkov and Florian Schäfer for useful discussions and review of this manuscript. 

This work has received funding from the European Union’s (EU) Horizon 2020 research and innovation program under Grant Agreement No. 820404 (iqClock project) and No. 860579 (MoSaiQC project). It also received funding from the European Partnership on Metrology, co-financed from the European Union’s Horizon Europe Research and Innovation Programme and by the Participating States, under project 22IEM01 TOCK. It further received funding from the Dutch National Growth Fund (NGF), as part of the Quantum Delta NL programme.

The raw data and the analysis tools used in this manuscript can be found in Reference \cite{heizenreder2025data}.
\end{acknowledgments}

%%%%%%%%%%%%%%%%%%%%%%%%%%%%%%%%%%%%%%%%%%%%%%%%%%%%%%%%%%%%%%%%%%%%%%%%%%%%%%%%%%%%%%%%%%%%%%%

\appendix
\section{Supplementary Information}
\label{App:a}
\subsection{Frequency Transfer}
\label{App:Frequency Transfer}
The stability of the frequency generated by RF TUNE is transferred to the optical domain via an Erbium fiber-based optical frequency comb optimized to have sub-kHz linewidth at the 689-nm strontium laser cooling transition (\({}^1S_0 \rightarrow {}^3P_1\)). Specifically, the repetition rate (\(f_{\mathrm{rep}}\)) of the comb is phase-locked to the tunable RF reference. In addition, the carrier-envelope offset frequency (\(f_{\mathrm{ceo}}\)), measured with a fiber-based \(f\)-to-\(2f\) interferometer, is locked to a fixed \(10~\mathrm{MHz}\) RF reference.
Typical state-of-the-art fiber combs exhibit linewidths on the order of tens of kilohertz~\cite{newbury2006reducing}, which is insufficient for strontium laser cooling at the targeted transition. 
Our Er-fiber oscillator was engineered such that both \(f_{\mathrm{rep}}\) and \(f_{\mathrm{ceo}}\) are minimally sensitive to pump power fluctuations, reaching sub-kHz optical linewidth~\cite{hutter2023femtosecond}. It has been shown that the linewidth of comb teeth reaches a minimum at a specific spectral position, referred to as the \emph{fixed point} ~\cite{washburn2005response}. Using our scheme, this fixed point can be positioned nearly arbitrarily. In our setup, the pump-induced fixed point was chosen to achieve a minimum linewidth of \(700~\mathrm{Hz}\) at \(689~\mathrm{nm}\).
A CW diode laser (Toptica DL pro) was locked to this narrow comb line and used as the MOT laser. By tuning the RF reference, the spectral position of the CW laser can be precisely adjusted.

\subsection{RF-Tuning}
\label{App:RF-Tuning}
\begin{figure}[]
  \centering
  \includegraphics[width=8.3cm]{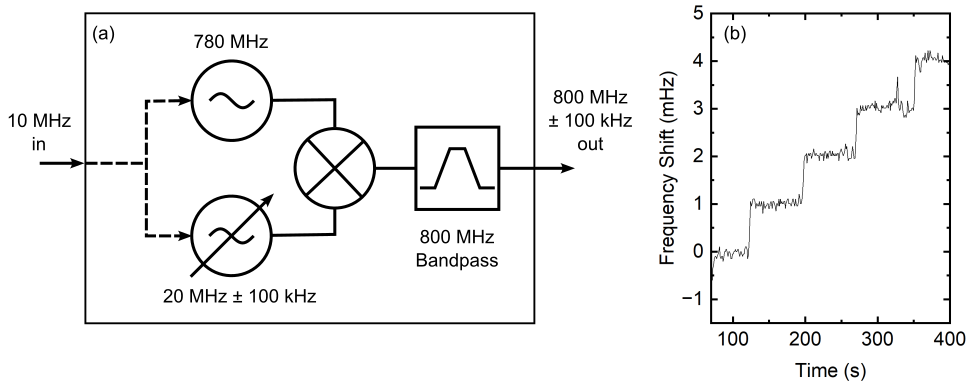}
  \caption[Tuneable RF-Source]{(a) Schematic of the tuneable
    \(800~\mathrm{MHz}\) source. Two quartz-oscillators (\(780~\mathrm{MHz}\),
    digitally tuneable \(20~\mathrm{MHz}\)) are mixed and bandpassed to generate
    the \(800~\mathrm{MHz}\) output. The optional \(10~\mathrm{MHz}\) reference
    was not connected for both oscillators. (b) Frequency shift steps at
    \(800~\mathrm{MHz}\) recorded with a counter while tuning the RF-source
    manually.}
  \label{fig:Tref}
\end{figure}
The repetition rate $f_{\mathrm{rep}}$ is stabilized to a tuneable RF source
(Toptica DFC RF TUNE). The schematic of this RF source is shown in
Fig.~\ref{fig:Tref}(a). A fixed frequency OCXO (oven controlled crystal
oscillator) at \(780~\mathrm{MHz}\) is mixed with a second digitally tuneable
signal at \(20~\mathrm{MHz}\). The resulting \(800~\mathrm{MHz}\) output is
tuneable within a bandwidth of \(\pm 100~\mathrm{kHz}\). The resolution of the
tuneable RF source was measured by counting the output mixed down to
\(20~\mathrm{MHz}\) with a GPS referenced K+K counter. As shown in
Fig.~\ref{fig:Tref}(b), the achievable frequency steps are 
\(1~\mathrm{mHz}\) or less.  The device was used in free running mode i.e. without the optional \(10~\mathrm{MHz}\) reference input. The overlapping Allan deviation of the free running tuneable RF-source is depicted in Fig.~\ref{fig:allan_variance_RF} (orange diamonds).

\subsection{Temperature Limit in a Five-Beam MOT}
\label{App:Temperature Limit}
The capture velocity of a standard six-beam magneto-optical trap (MOT) can be estimated by calculating the maximum velocity that can be slowed within the MOT beam profile. Assuming photon scattering at the maximum rate \( \gamma/2 \), where \( \gamma = \gamma_0 \sqrt{1+s} \), over the entire beam diameter \( D \), the capture velocity is given by
\begin{equation}
    v_{\text{6,cap}} = \sqrt{\frac{\hbar k \gamma}{m} D},
\end{equation}
where \( k = 2\pi/\lambda \) is the wave number and \( m \) the atomic mass~\cite{EnhancingCaptureVelocityDyMOT}.

In a five-beam MOT, gravitational acceleration limits the capture dynamics. We conservatively estimate the capture velocity by requiring that atoms be slowed by gravity within the horizontal beam diameter \( D \), yielding
\begin{equation}
    v_{\text{5,cap}} = \sqrt{2gD},
\end{equation}
provided that \( v_{\text{6,cap}} > v_{\text{5,cap}} \). Experimental results~\cite{Rodrigo_MOT_2021} show that capture velocities can easily exceed \( v_{\text{5,cap}} \) by an order of magnitude when atoms enter the MOT region counter-propagating the bottom beam (i.e., beam against gravity direction). In this regime, the bottom beam effectively acts as a white-light and Zeeman slower.
In steady state, the MOT temperature determines the typical thermal velocity, \( v_{\text{rms}} = \sqrt{3k_B T/m} \), which must remain well below \( v_{\text{5,cap}} \) to ensure confinement. This leads to the condition
\begin{equation}
    \frac{gmD}{k_B} \gg T_{\text{Doppler}} = \frac{\hbar \gamma}{2k_B},
\end{equation}
where \( T_{\text{Doppler}} \) is the Doppler cooling limit. 
If this condition is satisfied, a stable MOT can be formed in the five-beam configuration.

\subsection{Low Saturation Limit for the Broadband MOT}
\label{App:Saturation Limit}
In steady state, for the broadband MOT (see Fig.~\ref{fig:multi_vs_SF}), Eq.~\ref{eq:broad_modulated} is based on the condition:
\begin{equation}
    \frac{F_{\text{total}}}{m} = 0 = g - \frac{\hbar k \gamma_0 }{2m} \sum_{i=0}^{n} \frac{s_i}{1 + s_i + \left(2\delta_i/\gamma_0\right)^2},
\end{equation}
where \( \hbar k \gamma_0 / (2m) \approx 15.6g \). Each term in the sum is positive, since
\[
\frac{s_i}{1 + s_i + \left(2\delta_i/\gamma_0\right)^2} > 0,
\]
where \( s_i = I_i/I_0 \) is the saturation parameter of the \(i^\text{th}\) comb line, defined in terms of its specific intensity \( I_i \). 
To balance the gravitational force, the combined scattering force must approach \( g \), which requires that each term in the sum be small. This leads to the condition
\[
\alpha_i I_{\text{total}} / I_0 \ll \left(2\delta_i/\gamma_0\right)^2,
\]
where \( \alpha_i \) accounts for the power fraction of the total intensity in the \(i^\text{th}\) comb line. This inequality defines the low-saturation (low-intensity) limit, where coherent effects can be neglected, justifying the use of the simplified expression in Eq.~\ref{eq:broad_modulated}.

\subsection{Shift Measurements for Error Budget}
\label{App:Shifts Measurements}

To compute the error budget presented in Table~\ref{table:ErrorBudget}, we evaluate the sensitivity of the MOT position to the three primary sources of noise relevant for continuous BB MOT position spectroscopy at a magnetic field gradient of \(0.39\,\mathrm{G/cm}\): bias magnetic field fluctuations, atom number fluctuations, and intensity noise in the MOT beams.
Note that we assume very high polarization purity in our MOT beams, as we use a cleaning cube after the fiber launcher, followed by a quarter-wave plate to control the polarization. Therefore, polarization fluctuations before the cube are translated into power fluctuations of the MOT beam, and we do not consider polarization fluctuations further.

We begin with magnetic field fluctuations. It is important to distinguish between fluctuations in the magnetic field gradient and fluctuations in the bias magnetic field. Fluctuations in the gradient can be suppressed to first order by operating at the intersection point of different magnetic field gradient profiles, which depends on the bias field in gravity direction. At this point, small changes in the gradient affect only the frequency-to-position conversion factor, but do not significantly shift the MOT position. Therefore, we assume that a gradient stability of 1\% is sufficient. 

In contrast, fluctuations in the bias magnetic field are the dominant source of position uncertainty in our system.
To quantify the influence of the bias magnetic field, we use dedicated bias coils along each Cartesian direction (x, y, z) and measure the resulting change in the MOT’s position along the vertical (gravity) axis. To determine the MOT position, we use blue absorption imaging and extract the center position via a two-dimensional Gaussian fit. By precisely measuring the applied current and knowing the coil geometry, we calculate the expected magnetic field at the MOT location.
\begin{figure}[]
    \centering
    \includegraphics[width=8.3cm]{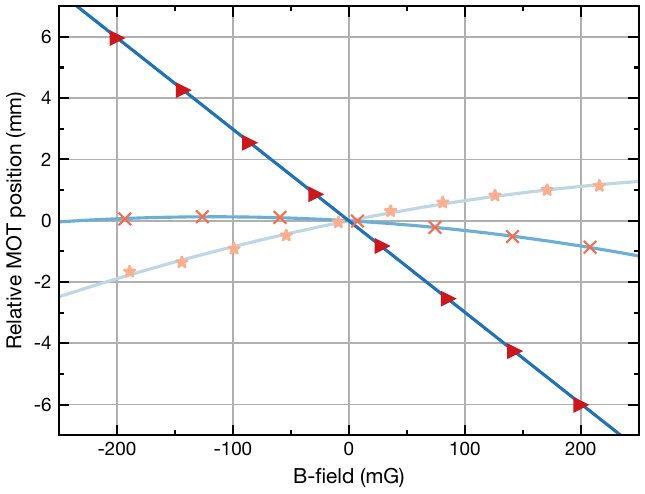}
    \caption[Bias Magnetic Field]{
    \label{fig:Bias mag field}
    Sensitivity of the MOT position to bias magnetic fields. The position shift along the gravity (z) axis is measured while scanning bias coil currents along the x, y, and z directions, with a magnetic field gradient of \(0.39~\mathrm{G/cm}\). A linear shift of \(29.9(2)~\mu\mathrm{m}/\mathrm{mG}\) is observed for fields along the z-direction (red triangles). For fields along the x (orange stars) and y (light red crosses) directions, the position shift follows a quadratic dependence \(z(B) = A \times (B - B_{\text{offset}})^2\), with fitted amplitudes \(A = -0.0096(4)\) and \(-0.0095(2)~\mu\mathrm{m}/\mathrm{mG}^2\), and offsets \(B_{\text{offset}} = 391.9(1.6)\) and \(-117.2(0.2)~\mathrm{mG}\)}, respectively.
\end{figure}
As shown in Fig.~\ref{fig:Bias mag field}, we observe the strongest effect for a magnetic field applied along the vertical direction. Here, we measure a linear shift of \(29.9(2)~\mu\mathrm{m}/\mathrm{mG}\) over a 400\,mG range (red triangles).
In contrast, applying a bias magnetic field along the x- and y-directions leads to a more subtle effect. While one might not expect a change in the MOT’s vertical position from horizontal fields, lateral displacement causes the atoms to experience different intensity regions of the bottom MOT beam due to its Gaussian intensity profile. This alters the local radiation pressure, requiring the MOT to shift vertically to reach a new equilibrium. The resulting position shift is well described by a quadratic dependence of the form \(z(B) = A \times (B - B_{\text{offset}})^2\). We extract similar amplitudes in both directions: \(A = -0.0096(4)~\mu\mathrm{m}/\mathrm{mG}^2\) in x, and \(-0.0095(2)~\mu\mathrm{m}/\mathrm{mG}^2\) in y. This agreement is consistent with the symmetry of the Gaussian beam profile in the transverse plane. However, the fitted offsets differ significantly: \(B_{\text{offset}} = 391.9(1.6)~\mathrm{mG}\) in x, and \(-117.2(0.2)~\mathrm{mG}\) in y. These differences suggest that the center of the beam profile is not perfectly aligned with the quadrupole magnetic field center. Centering would improve stability by rendering vertical MOT position changes only quadratically dependent on horizontal field changes.
Overall, the vertical (z-direction) bias magnetic field induces the largest shift, dominating the magnetic contribution to the systematic error. Shifts induced by bias fields in the transverse directions are more than an order of magnitude smaller, though still measurable and relevant for high-precision applications.

The second noise source we investigate is fluctuations in atom number. These are caused by small variations in oven temperature, beam pointing instabilities (particularly in the transverse cooling and Zeeman slower beams), and power fluctuations in the transverse cooling, Zeeman slower, and 2D blue MOT beams. Together, these factors affect the atom flux into the BB-MOT and are difficult to control precisely over long timescales.
To quantify the influence of atom number fluctuations, we operate the MOT in pulsed mode: the BB MOT is loaded for a fixed time, then the 2D blue MOT is switched off to isolate the BB MOT. After 100\,ms of wait, allowing the BB MOT to reach equilibrium, we measure the BB MOT position and atom number using blue absorption imaging.
\begin{figure}[]
    \centering
    \includegraphics[width=8.3cm]{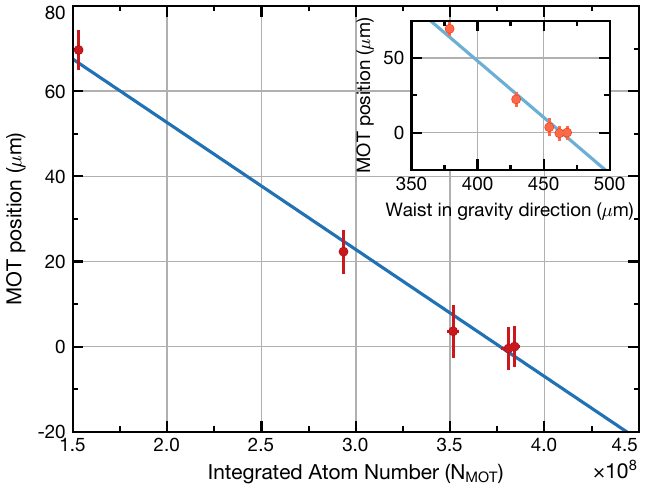}
    \caption[MOT Atom Number]{
    \label{fig:MOT Atom number}
    Sensitivity of the MOT position to atom number changes. The atom number is varied by loading the MOT for a finite time before imaging. All measurements are performed at a magnetic field gradient of \(0.33~\mathrm{G/cm}\). The main plot shows a linear dependence of the MOT position on atom number with a slope of \(29.8(2.7)~\mu\mathrm{m}/(10^8~\text{atoms})\). Inset: correlation between the in situ MOT cloud waist and vertical position, used to infer atom number changes during continuous operation. A linear fit yields a slope of \(-0.76(8)\).
    }
\end{figure}
Figure~\ref{fig:MOT Atom number} shows that increasing atom number results in a downward shift of the MOT’s equilibrium position. We observe a linear sensitivity of \(29.8(2.7)~\mu\mathrm{m}/(10^8~\text{atoms})\). This behavior is attributed to increased optical density at higher atom numbers, which leads to partial shielding of the atomic cloud from the vertical MOT beam. As a result, the radiation pressure is reduced, and the MOT shifts downward to a new equilibrium position where the local Zeeman shift compensates the force imbalance.
Since precise control of atom number is challenging in our system, we developed a method to infer and compensate for atom number fluctuations during continuous operation. 
Specifically, we use the vertical MOT waist size measured from absorption images as a proxy for the MOT atom number. We confirm that this size also correlates with the MOT position (see inset of Fig.~\ref{fig:MOT Atom number}), and we assume a similar correlation holds for fluorescence imaging. We propose this method instead of relying on total fluorescence counts, since at the high optical densities found in our BB MOT such counts lead to inaccurate atom number estimates. The measured sensitivity is \(-0.76(8)\). This technique therefore provides a means to correct for atom number fluctuations, relaxing the stability requirement on atom number by approximately an order of magnitude.

The third noise source we examine is intensity fluctuations in the MOT beams. Here, it is important to distinguish between fluctuations in the horizontal beams and in the single vertical beam. To characterize their respective influence on MOT position, we again use blue absorption imaging and measure the MOT's vertical displacement while varying the power of the two types of beams.
All measurements are performed at a magnetic field gradient of \(0.33~\mathrm{G/cm}\). Starting with the vertical beam, we vary its power from approximately 80\% to 100\% of its nominal value and observe a positive linear dependence of the MOT position on beam power. The measured slope is \(7.3(6)~\mu\mathrm{m}/\%\). This behavior is consistent with expectations: an increase in vertical beam power leads to a stronger radiation pressure from below. To restore force balance, the MOT shifts upward, requiring a smaller local Zeeman shift to reach equilibrium.
In contrast, fluctuations in the horizontal beam power result in a negative position shift, with a slope of \(-1.4(2)~\mu\mathrm{m}/\%\). This effect arises from increased compression of the MOT at higher horizontal beam intensities, which in turn increases atomic density. The resulting increase in optical density leads to partial shielding of the vertical beam, reducing the effective light force and requiring the MOT to shift downward to reestablish equilibrium.
Both vertical and horizontal power fluctuations could, in the future, be stabilized below the 1\% level using active power control, making their contribution to long-term position drift manageable.

\begin{figure}[H]
    \centering
    \includegraphics[width=8.3cm]{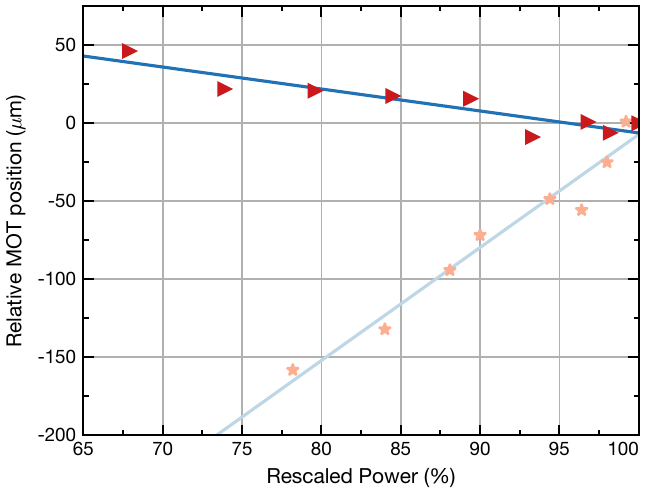}
    \caption[MOT Power Sensitivity]{
    \label{fig:MOT power sensitivity}
    Sensitivity of the MOT position to power fluctuations in the horizontal and vertical MOT beams. The position shift along the vertical (gravity) direction is measured while changing the power of each type of MOT beam. All measurements are taken at a magnetic field gradient of \(0.33~\mathrm{G/cm}\). For the horizontal beams (red triangles), we observe a linear shift of \(-1.4(2)~\mu\mathrm{m}/\%\), while for the vertical beam (orange stars), the shift is \(7.3(6)~\mu\mathrm{m}/\%\).
    }
\end{figure}

%%%%%%%%%%%%%%%%%%%%%%%%%%%%%%%%%%%%%%%%%%%%%%%%%%%%%%%%%%%%%%%%%%%%%%%%%%%%%%%%%%%%%%
\bibliography{Review_1/mainV2}
\end{document}